\documentclass[10pt,conference]{IEEEtran}
\IEEEoverridecommandlockouts
\usepackage{cite}
\usepackage{amsmath,amssymb,amsfonts}
\usepackage{graphicx}
\usepackage{textcomp}
\usepackage{xcolor}
\usepackage{empheq}
\usepackage{algorithmicx}
\usepackage{algorithm,algpseudocode}
\usepackage{algorithmicx}
\usepackage{setspace}
\usepackage{subcaption}

\usepackage{float}
\usepackage{color}
\usepackage{url}
\usepackage{dblfloatfix} 
\usepackage{graphicx}

\renewcommand{\baselinestretch}{0.98}

\usepackage{multirow}
\usepackage[short]{optidef}
\usepackage{balance}
\usepackage{hyperref}
\hypersetup{
    colorlinks=true,
    linkcolor=blue,
    filecolor=magenta,      
    urlcolor=cyan,
}

\def\BibTeX{{\rm B\kern-.05em{\sc i\kern-.025em b}\kern-.08em
    T\kern-.1667em\lower.7ex\hbox{E}\kern-.125emX}}

\begin{document}

\title{Optimal Design of Energy-Harvesting Hybrid VLC/RF Networks
\\\thanks{This work is supported in part by NSF under the grant CNS-1910153.}
}

\author{\IEEEauthorblockN{Amir Hossein Fahim Raouf\IEEEauthorrefmark{1},
Chethan Kumar Anjinappa\IEEEauthorrefmark{2}, and
Ismail Guvenc\IEEEauthorrefmark{1}}
\IEEEauthorblockA{\IEEEauthorrefmark{1}Department of Electrical and Computer Engineering,
North Carolina State University, Raleigh, NC}
\IEEEauthorblockA{\IEEEauthorrefmark{2}Ericsson Research, Santa Clara, CA 95054, USA}
\IEEEauthorblockA{amirh.fraouf@ieee.org, chethan.anjinappa@ericsson.com, iguvenc@ncsu.edu}
}

\maketitle

\begin{abstract}
In this paper, we consider an indoor downlink dual-hop hybrid visible light communication (VLC)/radio frequency (RF) scenario. For each transmission block, we dynamically allocate a portion of time resources to VLC and the other portion to RF transmission. In the first phase (i.e., VLC transmission), the LED carries both data and energy to an energy harvester relay node. In the second phase (i.e., RF communication), the relay utilizes the harvested energy to re-transmit the decoded information to the far RF user. During this phase, the LED continues to transmit power (no information) to the relay node, aiming to harvest energy that can be used in the next transmission block. We formulate the optimization problem in the sense of maximizing the data rate under the assumption of decode-and-forward (DF) relaying. As the design parameters, the direct current (DC) bias and the assigned time duration for VLC transmission are taken into account. In particular, the joint non-convex optimization is split into two sub-problems, which are then cyclically solved. In the first sub-problem, we fix the assigned time duration to VLC link and utilize the majorization-minimization (MM) procedure to solve the non-convex DC bias problem. 
In the second sub-problem, we fix the DC bias obtained in the previous step and solve the optimization problem for the assigned VLC link time duration.
Our results demonstrate that a higher data rate can be achieved by solving the joint problem of DC bias and time duration compared to solely optimizing the DC bias.
\end{abstract}


\section{Introduction}
Visible light communications~(VLC) is a promising complementary technology to radio-frequency~(RF) based wireless systems, as it can be used to offload users from~RF bands for releasing spectrum resources for other users while also simultaneously providing illumination~\cite{burchardt2014vlc}. Recently, hybrid~VLC-~RF systems are receiving attention in the literature, which can be designed to take advantage of both technologies, e.g., high-speed data transmission by VLC links while achieving seamless coverage through RF links~\cite{basnayaka2015hybrid}. In particular, a hybrid~VLC-RF system is desirable for indoor applications such as the Internet of things~(IoT) or wireless sensor networks~\cite{delgado2020hybrid}. Given that power constraint is a key performance bottleneck in such networks, a potential approach is to scavenge energy from the surrounding environment through energy harvesting~(EH). 

There have been some recent studies on enabling EH for a dual-hop hybrid VLC-RF communication system where the relay can harness energy from a VLC link (first hop), for retransmitting the data to the end user over the RF link (second hop).
For example, an optimal design in the sense of maximizing the data rate with respect to the direct current~(DC) bias by allocating an equal time portion for VLC and RF transmission is introduced in~\cite{rakia2016optimal}. In another work, both energy and spectral efficiency are investigated in~\cite{yapici2020energy} by taking into account the power consumption of the light emitting diodes (LEDs), which clearly shows the trade-off between energy and spectral efficiency and the importance of optimizing the DC bias. 
In \cite{zhang2021slipt}, the relay is chosen among multiple IoT devices which are randomly distributed within the coverage of the source. Based on the channel state information, an analytical approach is used to determine the end-to-end outage probability for two different transmission schemes.
In~\cite{peng2020performance}, Peng \textit{et al.} consider a mobile relay to facilitate the communications between the source and destination. The end-to-end outage probability of the system is analytically obtained and compared with the simulation results. They further extend this work and study the minimization problem of end-to-end outage probability under the constraints of both the average and the peak powers of the LED source in~\cite{peng2021end}.
In~\cite{tran2019ultra}, a hybrid VLC-RF ultra-small network is introduced where the optical transmitters deliver both the lightwave information and energy signals whereas a multiple-antenna RF access point (AP) is employed to transfer wireless power over RF signals.

\begin{figure}[t]
\centering
\includegraphics[trim=0.1cm 0.1cm 0.1cm 0.3cm, clip,width=\linewidth]{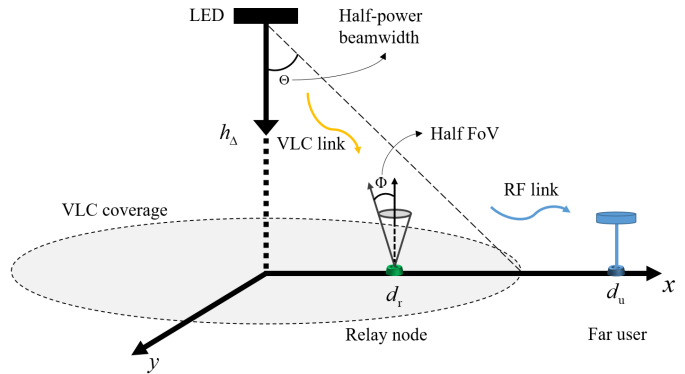}
\caption{The system model for the considered VLC-RF transmission scenario. The VLC link carries both data and energy to the relay node. The harvested energy is then used at the relay node to forward the data to the far RF user.}
\label{fig:system_model}
\end{figure}

The current literature on hybrid VLC-RF systems that also use EH is mainly limited to finding the optimal value of DC bias to either maximize the data rate or minimize the outage probability~\cite{rakia2016optimal,yapici2020energy,peng2020performance}. To our best knowledge, optimization of VLC and RF resources for hybrid RF/VLC links for a scenario such as in Fig.~\ref{fig:system_model} is not studied in the literature. In this paper, we investigate the performance of EH for an indoor hybrid VLC-RF scenario as in Fig.~\ref{fig:system_model}. We dynamically allocate a portion of each transmission block to VLC and the rest to RF transmission. The LED transmits both data and energy to a relay node with energy harvesting capability in the first phase as illustrated in  Fig.~\ref{fig:transmission_block} (i.e., VLC transmission). During the second phase (RF communication), the relay transmits the decoded information to the distant RF user using the harvested energy. Also, during this phase, the~LED continues to transmit power (no information) to the relay node, aiming to harvest energy that can be utilized by the~RF relay in the next transmission block.

For this specific scenario, we formulate an optimization problem for maximizing the data rate at the far user. In particular, we consider both the DC bias and the assigned time duration to VLC link as the design parameters. We split the joint non-convex optimization problem into two sub-problems and cyclically solve them. First, we fix the assigned time duration for VLC transmission and solve the non-convex problem for DC bias by employing the majorization-minimization~(MM) procedure~\cite{Yusuf_VLC,sun2016majorization}. Second, we fix the DC bias obtained from the previous step and solve the optimization problem for the assigned time duration to~VLC~link.

The remainder of this paper is organized as follows. In Section \ref{sec:sys_model}, we describe our system model. The end-to-~end data rate is formulated in Section~\ref{sec:per_analysis} along with the respective optimization problem and our approach to solve the optimization problem.
The numerical results are presented in Section~\ref{sec:num_results}, and finally, the paper concludes with Section~\ref{sec:conclusion}.

\section{System Model}\label{sec:sys_model}
\begin{figure}[t]
\centering
\includegraphics[width=\linewidth]{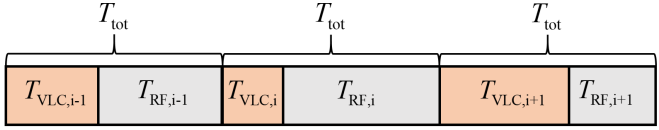}
\caption{The transmission block under consideration with consecutive time periods dedicated for VLC and RF links.}
\label{fig:transmission_block}
\end{figure}

Fig.~\ref{fig:system_model} illustrates the hybrid VLC-RF system under consideration. We assume a relay equipped with a single photo-detector~(PD), energy-harvesting circuity, and a transmit antenna for RF communications. Let $T_{{\rm tot},{\rm i}}$ denote the~$i^{\rm{th}}$ block transmission time which is measured in seconds. Also,~$\tau_{\rm i}$~(unitless value) is the portion of time that is allocated to transmit information and energy to the relay node in the $i^{\rm{th}}$ time block. Thus, the duration of this phase in seconds is $T_{\rm{VLC},{\rm i}} = \tau_{\rm i} T_{\rm{tot}}^{\rm (i)}$. We assume that the block transmission time is constant. Hereafter, we drop the superscript of $T_{\rm{tot}}^{\rm (i)}$ in the sequel to simplify notation. Fig.~\ref{fig:transmission_block}
depicts the transmission block under consideration. Without loss of generality, we assume that $T_{\rm{tot}} = 1$ second. 

\subsection{VLC Link}
In the first hop, the LED transmits both energy and information to the relay node through the VLC link. The non-negativity of the transmitted optical signal can be assured by adding DC bias (i.e., ${I_{\rm{b},{\rm i}}}$) to the modulated signal, i.e., ${{x_{\rm{t},{\rm i}}}\left( t \right) = P_{\rm{LED}}\big( {{x_{\rm{s},{\rm i}}}\left( t \right) + {I_{{\rm{b}},{\rm i}}}} \big)}$ where $P_{\rm{LED}}$ denotes the LED power per unit (in W/A) and $x_{\rm{s},{\rm i}}$ is the modulated electrical signal. We assume that the information-bearing signal is zero-mean, and satisfies the peak-intensity constraint of the optical channel such that~\cite{1}
\begin{equation}\label{eq:1}
A_{{\rm i}} \le \min \left( {{I_{\rm{b},{\rm i}}} - {I_{\rm{L}}},{I_{\rm{H}}} - {I_{\rm{b},{\rm i}}}} \right),
\end{equation}
where $A_{{\rm i}}$ denotes the peak amplitude of the input electrical signal (i.e., $\max (\left| {{x_{\rm{s},{\rm i}}}(t)} \right|) = A_{{\rm i}}$), and ${I_{\rm{b},{\rm i}}} \in \left[ {{I_{\rm{L}}},{I_{\rm{H}}}} \right]$ with ${I_{\rm{H}}}$ and ${I_{\rm{L}}}$ being the maximum and minimum input currents of the DC offset, respectively. 
Let ${B_{{\rm{VLC}}}}$ denote the double-sided signal bandwidth. 

Then, the information rate associated with the optical link between the AP and relay node within a block with $T_{\rm{tot}} = 1$ second, is given as~\cite{1}
\begin{equation}\label{eq:2}
{R_{{\rm{VLC}},{\rm i}}} = T_{\rm{VLC},{\rm i}}{B_{{\rm{VLC}}}}{\log _2}\left( {1 + \frac{e}{{2\pi }}\frac{{{{\left( {\eta P_{\rm{LED}}A_{{\rm i}}{H_{\rm{VLC}}}} \right)}^2}}}{{\sigma _{{\rm{VLC}}}^2}}} \right),
\end{equation}
where $\eta$ is the photo-detector responsivity in A/W and ${H_{{\rm{VLC}}}}$ is the optical DC channel gain. In \eqref{eq:2}, $\sigma _{{\rm{VLC}}}^2$ is the power of shot noise at the PD which is given as $\sigma _{{\rm{VLC}}}^2 = {q_{\rm{e}}}{I_{\rm{i}}}{B_{{\rm{VLC}}}}$ where ${q_{\rm{e}}}$ is the charge of an electron and ${I_{\rm{i}}}$ is the induced current due to the ambient light. The optical DC channel gain of the VLC link can be written as
\begin{equation}\label{eq:3}
{H_{{\rm{VLC}}}} = \frac{{\left( {m + 1} \right){A_{\rm{p}}}}}{{2\pi \left( {h_\Delta ^2 + d_{\rm{r}}^2} \right)}}{\cos ^m}\left( {{\phi _{\rm{r}}}} \right)\cos \left( {{\theta _{\rm{r}}}} \right)\Pi \left( {\left| {{\theta _{\rm{r}}}} \right|,\Phi } \right),
\end{equation}
where ${h_\Delta }$ and ${d_r}$ are the vertical and horizontal distances, respectively, between AP and the relay node, and ${\phi _r}$ and ${\theta _r}$ are the respective angle of irradiance and incidence, respectively. The Lambertian order is $m = {{ - 1} \mathord{\left/{\vphantom {{ - 1} {{{\log }_2}\left( {\cos \left( \Theta  \right)} \right)}}} \right.
 \kern-\nulldelimiterspace} {{{\log }_2}\left( {\cos \left( \Theta  \right)} \right)}}$ where $\Theta $ is the half-power beamwidth of each LED, and ${A_{\rm{p}}}$ and $\Phi $ are the detection area and half field-of-view (FoV) of the PD, respectively. The function $\Pi \left( {x,y} \right)$ is 1 whenever $x \le y$, and is 0 otherwise. 
 We also assume that the relay node distance follows a Uniform distribution with $\mathcal{U}\left[ {d_{\rm{r},{\min }},d_{\rm{r},{\max }}} \right]$, and that relay PD is looking directly upward.

The harvested energy at this phase can be computed as
\begin{equation}\label{eq:4}
{E_{\rm{1},{\rm i}}} = 0.75T_{\rm{VLC},{\rm i}}{I_{{\rm{DC}},{\rm i}}}{V_{\rm{t}}}\ln \left( {1 + \frac{I_{{\rm{DC}},{\rm i}}}{{{I_0}}}} \right),
\end{equation}
where ${V_{\rm{t}}}$ is the thermal voltage, ${I_0}$ is the dark saturation current, and ${I_{{\rm{DC}},{\rm i}}}$ is the DC part of the output current given as 
$ I_{{\rm{DC}},{\rm i}} = \eta {H_{{\rm{VLC}}}}P_{\rm{LED}}{I_{\rm{b},{\rm i}}}$.
In the time period $T_{\rm{RF},{\rm i}} = 1-T_{\rm{VLC},{\rm i}}$, the aim is to maximize the harvested energy while the relay transmits the information to the far user over the RF link. Thus, during second phase the LED eliminates the alternating current (AC) part and maximizes the DC bias, i.e., $A_{{\rm i}} = 0$ and $I_{\rm{b},{\rm i}} =
I_{\rm{H}}$. Mathematically speaking, the harvested energy during the second phase can be expressed as
\begin{equation}\label{eq:5}
{E_{2,{\rm i}}} = 0.75T_{\rm{RF},{\rm i}}{I_{\rm DC, \rm max}}{V_{\rm{t}}}\ln \!\left(\! {1 + \frac{I_{\rm DC, \rm max}}{I_0}}\! \right),
\end{equation}
where $I_{\rm DC, max} = \eta {H_{{\rm{VLC}}}}P_{\rm{LED}}{I_{\rm{H}}}$.
The total harvested energy at the relay that can be utilized for transmitting the decoded symbol to the far user through an RF link can be calculated~as
\begin{equation}\label{eq:6}
\begin{split}
E_{\rm{h},{\rm i}} = &{E_{1,{\rm i}}} + {E_{2,{\rm i-1}}}\\
 = & f\Big(T_{\rm{VLC},{\rm i}}I_{\rm{b},{\rm i}}\ln{\big(1+\frac{\eta H_{\rm{VLC}} P_{\rm{LED}} I_{\rm{b},{\rm i}}}{I_0}\big)} \\ 
 & + T_{\rm{RF},{\rm i-1}}I_{\rm{H}}\ln{\big(1+\frac{\eta H_{\rm{VLC}} P_{\rm{LED}} I_{\rm{H}}}{I_0}\big)}\Big),
\end{split}
\end{equation}
where $f=0.75 \eta H_{\rm{VLC}} P_{\rm{LED}} V_{\rm{t}}$, ${E_{2,{\rm i}}}$ represents the harvested energy during the RF transmission in the previous transmission block. In this paper, we assume that the initial harvested energy is 0 (i.e., ${E_{2,{\rm 0}}} = 0$). 

As it can be readily checked from~\eqref{eq:1}, increasing $I_{\rm b, \rm i}$ leads to a decrease in $A_{\rm i}$ and, consequently, it decreases the information rate associated with the VLC link. On the other hand, decreasing $I_{\rm b, \rm i}$ limits the harvested energy that can be obtained during VLC transmission (i.e., $E_{\rm{1},{\rm i}}$). 

\subsection{RF Link}
In the second hop, the relay re-transmits the information to the far user through the RF link by utilizing the harvested energy. The relaying operation is of decode-and-forward (DF) type. We assumed that the user is off the AP horizontally by a distance ${d_{\rm{u}}}$, which follows a Uniform distribution with $\mathcal{U}\left[ {d_{\rm{u},{\rm min}},d_{\rm{u},{\rm max}}} \right]$. Let ${B_{{\rm{RF}}}}$ denote the bandwidth for the RF system and ${N_0}$ denote the noise power which can be defined as ${{N_0} = {P_0} + 10{\log _{10}}\left( {{B_{{\rm{RF}}}}} \right) + {N_{\rm{F}}}}$ where ${P_0}$ is the thermal noise power, and ${N_{\rm{F}}}$ is the noise figure. Further, assume that the relay re-transmits the electrical signal with normalized power. The respective information rate is given as 
\begin{equation}\label{eq:7}
{R_{\rm{RF},{\rm i}}} = T_{\rm{RF},{\rm i}}{B_{{\rm{RF}}}}{\log _2}\left( {1 + \frac{{{P_{\rm{h}}}{{\left| {{h_{{\rm{RF}}}}} \right|}^2}}}{{{G_{{\rm{RF}}}}{N_0}}}} \right),
\end{equation}
where ${h_{{\rm{RF}}}}$ denotes the Rayleigh channel coefficients, ${{P_{\rm{h},{\rm i}}} = {E_{\rm{h},{\rm i}}}/T_{\rm{RF},{\rm i}}}$ is the transmit power and ${G_{{\rm{RF}}}}$ is the path loss model for RF link and can be expressed as
\begin{equation}\label{eq:8}
{G_{{\rm{RF}}}} = {\left( {\frac{{4\pi {d_0}}}{\lambda }} \right)^2}{\left( {\frac{{{d_{{\rm{u}}}}}}{{{d_0}}}} \right)^\beta },
\end{equation}
where $\lambda $ being the used RF carrier wavelength, ${d_0} = 1$ m is the reference distance, and $\beta $ is the path loss exponent, which generally takes value between [1.6, 1.8]. 

The achievable information rate is limited by the smaller information rate between the VLC link and the RF link and can be expressed as~\cite{guo2021achievable}
\begin{equation}\label{eq:9}
    R_{\rm{VLC-RF},{\rm i}} = \min{(R_{\rm{VLC},{\rm i}}, R_{\rm{RF},{\rm i}})}.
\end{equation}

\section{Optimization Framework and Solution}\label{sec:per_analysis}

\begin{figure}[t]
\centering
\includegraphics[width=\linewidth]{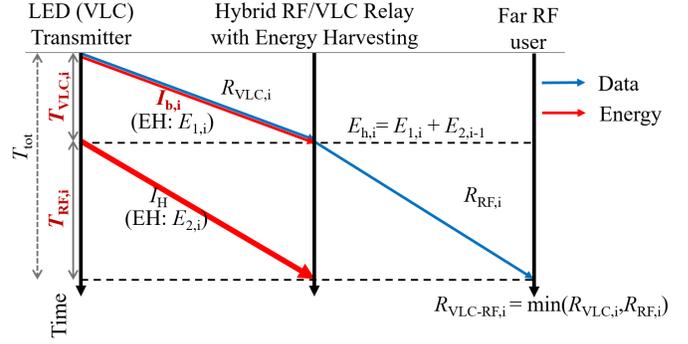}
\caption{Summary of the optimization problem. Here, we consider $I_{\rm b, \rm i}$, $T_{\rm VLC, \rm i}$, and $T_{\rm RF, \rm i}$ as the optimization variables to maximize the system data rate.}
\label{fig:schematic_sys}
\end{figure}
Fig. \ref{fig:schematic_sys} illustrates the summary of our optimization problem in which $I_{\rm b, \rm i}$, $T_{\rm VLC, \rm i}$ and $T_{\rm RF, \rm i}$ are considered as the optimization variables to maximize the system data rate. Recalling the information rate in VLC link (i.e., \eqref{eq:2}) and RF link (i.e., \eqref{eq:6}), the optimization problem can be written as
\begin{equation}\label{eq:10}
\begin{aligned}
\max_{I_{\rm{b},{\rm i}}, T_{\rm{VLC},{\rm i}}, T_{\rm{RF},{\rm i}}} \ & R_{\rm{VLC-RF},{\rm i}}\\
\textrm{s.t.} \quad & c_1: I_{\rm{L}} \leq I_{\rm{b},{\rm i}} \leq I_{\rm{H}}, \\
  & c_2:  T_{\rm{VLC},{\rm i}} \!+\! T_{\rm{RF},{\rm i}} \!=\! 1, T_{\rm{VLC},{\rm i}} >0, T_{\rm{RF},{\rm i}} > 0  \\
  & c_3:  R_{\rm{th}} \leq R_{\rm{RF}}^{\rm{i}}.
\end{aligned}
\end{equation}
where $R_{\rm{th}}$ is a predefined threshold value, constraints $c_1$ and $c_2$ are imposed to avoid any clipping and guarantee that the LED operates in its linear region and $c_3$ is added to satisfy the minimum required data rate. The above joint-optimization problem is non-smooth (due to the $\min$ operator) and non-convex (due to the objective function and constraint $c_3$). We reformulate the above optimization problem in the epigraph form to remove the non-smoothness in the objective function. The epigraph form of~\eqref{eq:10} can be written as
\begin{equation}\label{eq:11}
\begin{aligned}
&\max_{\phi, I_{\rm{b},{\rm i}}, T_{\rm{VLC},{\rm i}}, T_{\rm{RF},{\rm i}}} \quad  \phi \\
\textrm{s.t.} \quad & c_1^{\prime}, c_2, c_3,\\
\quad  & c_4:  \phi \leq R_{\rm{VLC},{\rm i}} \\
\quad  & c_5: \phi \leq R_{\rm{RF},{\rm i}}.
\end{aligned}
\end{equation}
The above equivalent optimization problem to \eqref{eq:10} solves the non-smoothness, while it is still non-convex. Let ${\alpha = {e\left( {\eta P_{\rm{LED}}{H_{\rm{VLC}}}} \right)}^2/(2\pi\sigma _{{\rm{VLC}}}^2)}$, ${\beta = \eta {H_{{\rm{VLC}}}}P_{\rm{LED}}}$, and
${\zeta = |h_{\rm{RF}}|^2/(G_{\rm{RF}}N_0)}$. Substituting~\eqref{eq:2} and~\eqref{eq:7} in~\eqref{eq:11}, we~have
\begin{equation}\label{eq:12}
\begin{aligned}
 &\max_{\phi, I_{\rm{b},{\rm i}}, A_{\rm i}, E_{\rm h,{\rm i}}, T_{\rm{VLC},{\rm i}},T_{\rm{RF},{\rm i}}} \  \phi \\
\textrm{s.t.} \quad & c_1, c_2, \\
\quad & c_3:  T_{\rm{RF},{\rm i}} B_{\rm{RF}}\! \log_2\!\!\left(\!\!1 \!+\! \frac{\zeta E_{\rm{h},{\rm i}}}{T_{\rm{RF},{\rm i}}}\!\right)\! \!\geq \! \! R_{\rm{th}},\\
\quad  & c_4: T_{\rm{VLC},{\rm i}} B_{\rm{VLC}} \log_2(1 + \alpha A_{\rm i}^2)\!\! \geq \! \phi ,\\
\quad  & c_5: T_{\rm{RF},{\rm i}} B_{\rm{RF}} \log_2\!\left(\!1 \! + \! \frac{\zeta E_{\rm{h},{\rm i}}}{T_{\rm{RF},{\rm i}}}\right) \! \geq \! \phi ,\\
\quad  & c_6:   \min \big( I_{\rm{b},{\rm i}} - I_{\rm{L}}, I_{\rm{H}} - I_{\rm{b},{\rm i}}\big) \! \geq \! A_{\rm{i}},\\
\quad  & c_7:  f\Big( T_{\rm{VLC},{\rm i}}I_{\rm{b},{\rm i}}\ln{\big(1 + \frac{\beta I_{\rm{b},{\rm i}}}{I_0}\big)} \\
\quad & \quad +\! T_{\rm{RF},{\rm i-1}} I_{\rm{H}}\! \ln\!{\big( 1\! + \!\frac{\beta I_{\rm{H}}}{I_0}\big)} \!\Big)\!\! \geq \! E_{\rm{h},{\rm i}}.
\end{aligned}
\end{equation}
In the above optimization problem,  $A_{\rm i}^2$ is used in $c_4$ and $c_6$ is still non-smooth. Here, we relax $c_6$ by using the Proposition 1 from~\cite{guo2021achievable}. Intuitively, the higher the term $I_{b,i}$ higher the energy harvesting, however, a negative effect on the rate beyond $(I_L + I_H)/2$. Thus, the optimal value of the term $I_{b,i}$ would be within $(I_L + I_H)/2$ and $I_H$ (and not the other regime $0 \leq I_{b,i} \leq I_H)$. The above restriction enforces $0 \leq A_{\rm i} \leq  I_{\rm{H}} - I_{\rm{b},{\rm i}}$ benefiting in getting rid of the non-smooth $\min$ operator (see~\eqref{eq:12}) as well. This leads to $c_1^{\prime}: (I_{\rm min} + I_{\rm max})/{2} \leq I_{\rm b, \rm i} \leq I_{\rm max}$ and $c_6^{\prime}: 0 \leq A_{\rm i} \leq  I_{\rm{H}} - I_{\rm{b},{\rm i}}$. The constraints $c_3, c_4, c_5, c_7$ are jointly non-convex. In this regard, we split the joint optimization problem into two sub-problems and solve them in a cyclic fashion.\looseness = -1


\subsection{Sub-problem 1} 
First, we fix $T_{\rm{VLC},{\rm i}}$ (and hence $T_{\rm{RF},{\rm i}} = 1- T_{\rm{VLC},{\rm i}}$)  and solve the maximization problem for $\phi$ over $I_{\rm{b},{\rm i}}, A_{\rm{i}}, E_{\rm{h},{\rm i}}$. Sub-problem 1 can be written as
\begin{equation}\label{eq:13}
\begin{aligned}
\max_{\phi, I_{\rm{b},{\rm i}},A_{\rm{i}},E_{\rm{h},{\rm i}} } \quad & \phi \\
\textrm{s.t.} \quad & c_1^{\prime}, c_3, c_4, c_5, c_7\\
\quad &  c_6^{\prime}:   0 \leq A_{\rm{i}} \leq I_{\rm{H}} - I_{\rm{b},{\rm i}},
\end{aligned}
\end{equation}
where the constraints $c_3, c_4, c_5$ are conditionally convex. 

\textbf{Assumption:}
The typical illumination requirement in an indoor VLC environment results in a high transmit optical intensity, which can provide a high signal-to-noise ratio (SNR) at the receiver~\cite{hanzo2012wireless, wang2018physical}. In this paper, we assume that SNR for the VLC link is much greater than 1 (in linear scale); i.e., $\alpha (A_{\rm{i}})^2 \gg 1$. In this condition, we further utilize ${\log(1+x) \approx \log(x)}$ in the constraints $c_4$. Thus, the optimization problem can be written~as
\begin{equation}\label{eq:14}
\begin{aligned}
\max_{\phi, I_{\rm{b},{\rm i}},A_{\rm{i}},E_{\rm{h},{\rm i}}} \quad  & \phi \\
\textrm{s.t.} \quad & c_1^{\prime}, c_3, c_5, c_6^{\prime}\\
\quad & c_4^{\prime}:  T_{\rm{VLC},{\rm i}} B_{\rm{VLC}} \log_2(\alpha A_{\rm{i}}^2) \geq \phi ,\\
\quad  & c_7^{\prime}:  f\bigg( T_{\rm{VLC},{\rm i}}I_{\rm{b},{\rm i}}\ln{\Big(\frac{\beta I_{\rm{b},{\rm i}}}{I_0}\Big)} \\
 &\quad +  \! T_{\rm{RF},{\rm i-1}} I_{\rm{H}}\ln\!{\Big(\! \frac{\beta I_{\rm{H}}}{I_0}\Big)} \bigg)\! \geq \! E_{\rm{h},{\rm i}}.
\end{aligned}
\end{equation}
In \eqref{eq:14}, $c_4^{\prime}$ is a convex constraint while $c_7^{\prime}$ is still not convex. We further utilize the first-order Taylor series and MM approach to relax this constraint~\cite{Yusuf_VLC,sun2016majorization}. As a result, $c_7^{\prime}$ can be replaced with 
\begin{equation}\label{eq:15}
c_7^{\star}: \quad  g(I_{\rm b,{\rm{i}}}) = g_0\big(I_{\rm{b},{\rm i}}(t)\big) + \frac{\partial g\big(I_{\rm{b},{\rm i}}(t)\big)}{\partial I_{\rm{b},{\rm i}}} \big(I_{\rm{b},{\rm i}} - I_{\rm{b},{\rm i}}(t)\big),
\end{equation}
where 
\begin{equation}\label{eq:16}
\begin{aligned}
g_0(I_{\rm{b},{\rm i}}(t)) =&  f\bigg(T_{\rm{VLC},{\rm i}} I_{\rm{b},{\rm i}}(t)\ln{\Big(\frac{\beta I_{\rm{b},{\rm i}}(t)}{I_0}\Big)}\\ & +
 T_{\rm{RF},{\rm i}-1}I_{\rm{H}}\ln{\Big(\frac{\beta I_{\rm{H}}}{I_0}\Big)}\bigg),
 \end{aligned}
 \end{equation}
and\\
 \begin{equation}\label{eq:17}
 \begin{aligned}
 \frac{\partial g(I_{\rm{b},{\rm i}}(t))}{\partial I_{\rm{b},{\rm i}}} & = f T_{\rm{VLC},{\rm i}}\bigg(\ln{\Big(\frac{\beta I_{\rm{b},{\rm i}}(t)}{I_0}\Big)} + \frac{\beta I_{\rm{b},{\rm i}}(t)}{I_0 + \beta I_{\rm{b},{\rm i}}(t)} \bigg).
 \end{aligned}
  \end{equation}
In \eqref{eq:15}, the term $t$ is an index-term and denotes the iteration index for the MM approach. The MM procedure on \eqref{eq:15} operates iteratively. We first solve the problem for some initial values of $I_{\rm b,{\rm i}}(t)$. Then, we update the value of $I_{\rm b,{\rm i}}(t)$ at each iteration until it remains the same for two consecutive iterations, or the change between two consecutive iterations is not appreciable.

Overall, the optimization sub-problem 1 is as follows:
\begin{equation}\label{eq:18}
\begin{aligned}
& \max_{\phi, I_{\rm{b},{\rm i}},A_{\rm{i}},E_{\rm{h},{\rm i}} } \quad  \phi \\
\textrm{s.t.} \quad & c_1, c_3, c_4^{\prime}, c_5, c_6^{\prime}, c_7^{\star}.
\end{aligned}
\end{equation}
We iteratively solve the above sub-problem 1 until its convergence. Once the above sub-problem converges, we continue with sub-problem 2 which is elaborated in the~following.

\subsection{Sub-problem 2}
In here, we fix $I_{\rm{b},{\rm i}}$ obtained from sub-problem 1 and solve the problem for maximizing $\phi$ over the variables $T_{\rm{VLC},{\rm i}}, T_{\rm{RF},{\rm i}},  E_{\rm{h},{\rm i}}$. The optimization problem can be expressed as
\begin{equation}\label{eq:19}
\begin{aligned}
& \max_{\phi, T_{\rm{VLC},{\rm i}}, T_{\rm{RF},{\rm i}},E_{\rm{h},{\rm i}} } \quad  \phi \\
\textrm{s.t.} \quad & c_2,  c_3,  c_4^{\prime},  c_5, c_7^{\prime}.
\end{aligned}
\end{equation}
In \eqref{eq:19}, the objective function and constraints $c_2, c_4^{\prime}, c_7^{\prime}$ are linear, whereas the constraints $c_3$ and $c_5$ are convex constraints which result in a convex optimization problem.

\begin{table}[t]
\caption{System and channel parameters}
\label{table1}
\begin{center}
\scalebox{0.9}{
\begin{tabular}{ |l|l| } 
 \hline
 \textbf{Parameter} & \textbf{Numerical Value} \\ \hline
User distance ($d_{\rm{u},{\rm{min}}}$,$d_{\rm{u},{\rm{max}}}$) &  [4,8] m  \\ \hline 
Relay distance ($d_{\rm{r},{\rm{min}}}$,$d_{\rm{r},{\rm{max}}}$) & [0,2] m  \\ \hline 
LED power ($P_{\rm{LED}}$) & 1.5 W/A \\ \hline
Noise figure ($N_{\rm{F}}$) & 9 dB  \cite{yapici2020energy} \\ \hline 
RF signal bandwidth $B_{\rm{RF}}$ &  10 MHz \cite{guo2021achievable} \\ \hline 
VLC signal bandwidth $B_{\rm{VLc}}$ &  10 MHz \cite{guo2021achievable} \\ \hline 
Thermal noise ($P_0$) & -174 dBm/Hz \cite{yapici2020energy} \\ \hline 
RF frequency ($f_{\rm{c}}$) &  \{2.4, 5\} GHz \cite{yapici2020energy} \\ \hline 
Minimum DC bias ($I_{\rm{min}}$) & 100 mA  \cite{yapici2020energy}\\ \hline
Maximum DC bias ($I_{\rm{max}}$) & 1 A  \cite{yapici2020energy} \\ \hline
Photo-detector responsivity ($\eta$ ) & 0.4 A/W \cite{1}\\ \hline
Thermal voltage ($V_{\rm{t}}$) & 25 mV \cite{yapici2020energy} \\ \hline
Dark saturation current ($I_0$) & $10^{-10}$ A \cite{yapici2020energy} \\ \hline
Half FoV ($\Phi$) & $60^{\circ}$ \cite{yapici2020energy} \\ \hline
Half-power beamwidth ($\Theta$) & $60^{\circ}$ \cite{yapici2020energy} \\ \hline
Electron charge ($q_{\rm{e}}$) & $1.6 \times 10^{-19}$ \\ \hline
Induced current ($I_{\rm{i}}$) & $5840 \times10^{-6}$ \cite{yapici2020energy} \\ \hline
PD detection area ($A_{\rm{p}}$) & $10^{-4}$ $\rm{m}^2$ \cite{yapici2020energy} \\ \hline
AP relative height ($h_{\Delta}$) & 2 m \cite{yapici2020energy} \\ \hline
Data rate threshold ($R_{\rm th}$) & $10^6$ b/s \\ \hline
\end{tabular}}
\end{center}
\end{table}

\section{Numerical Results}\label{sec:num_results}

In this section, we demonstrate the performance of hybrid VLC-RF scheme under consideration as in Fig.~\ref{fig:system_model}. For the convenience of the reader, the channel and system parameters are summarized in Table \ref{table1}. We consider four distinct cases:

\begin{figure}[t!]
\centering
\includegraphics[trim=0.2cm 0.1cm 0.2cm 0.3cm, clip,width=1.05\linewidth]{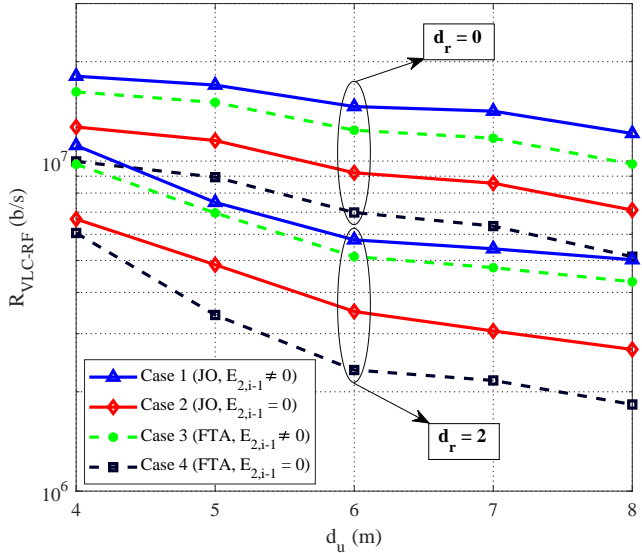}
\caption{The performance of optimal data rate for different user distances when the relay is located at $d_{\rm r} = 0$.}
\label{fig:rate_diff_user_dis}
\end{figure}

\begin{figure*}[!h]
\centering
\begin{subfigure}{0.32\linewidth} 
\centering
\includegraphics[trim=0.45cm 0.45cm 0.45cm 0.45cm, clip,width=\linewidth]{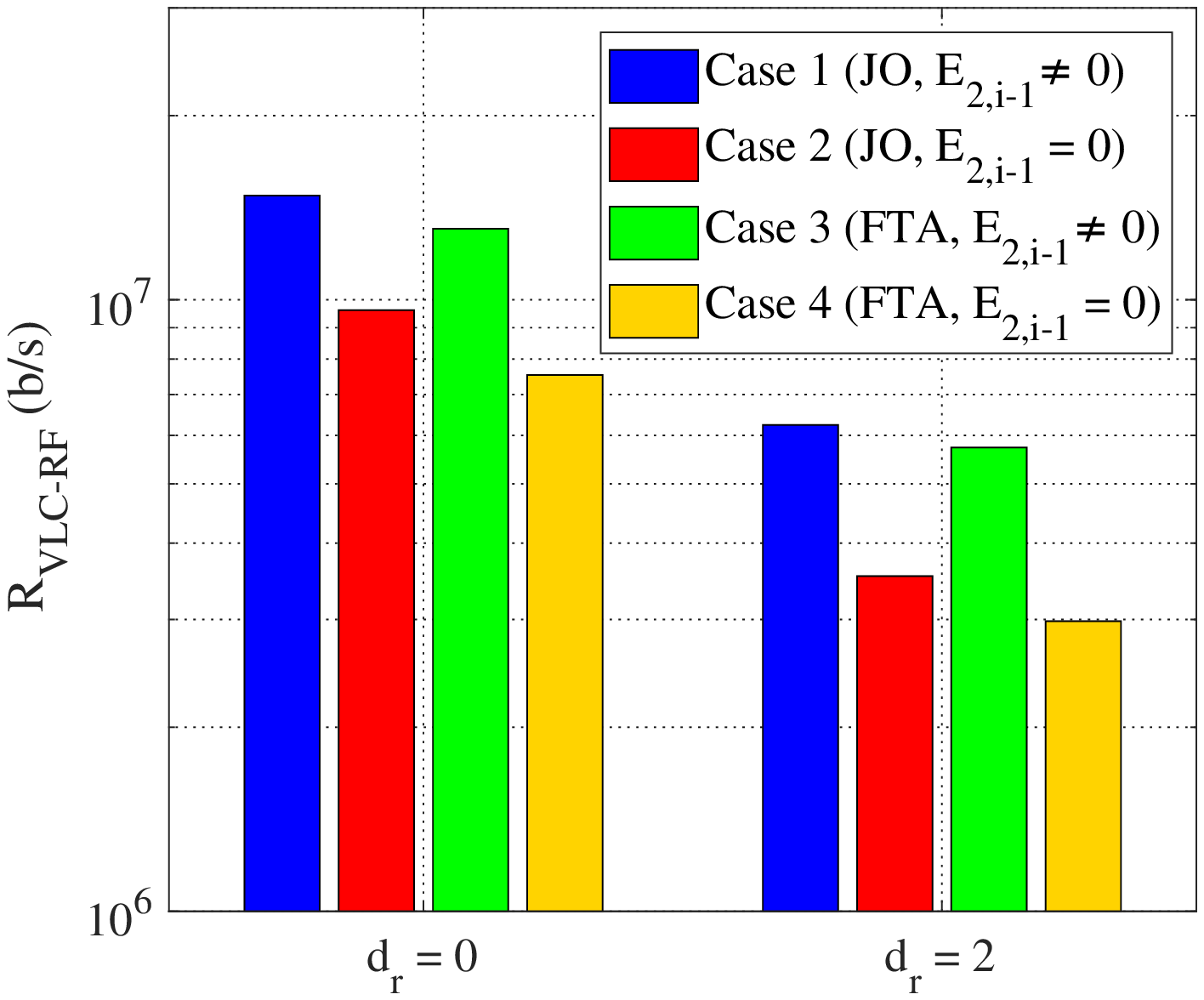} 
\caption{Optimal data rate}\label{fig:rate_random_2G} 
\end{subfigure}
\begin{subfigure}{0.32\linewidth} 
\centering
\includegraphics[trim=0.3cm 0.45cm 0.45cm 0.45cm, clip,width=\linewidth]{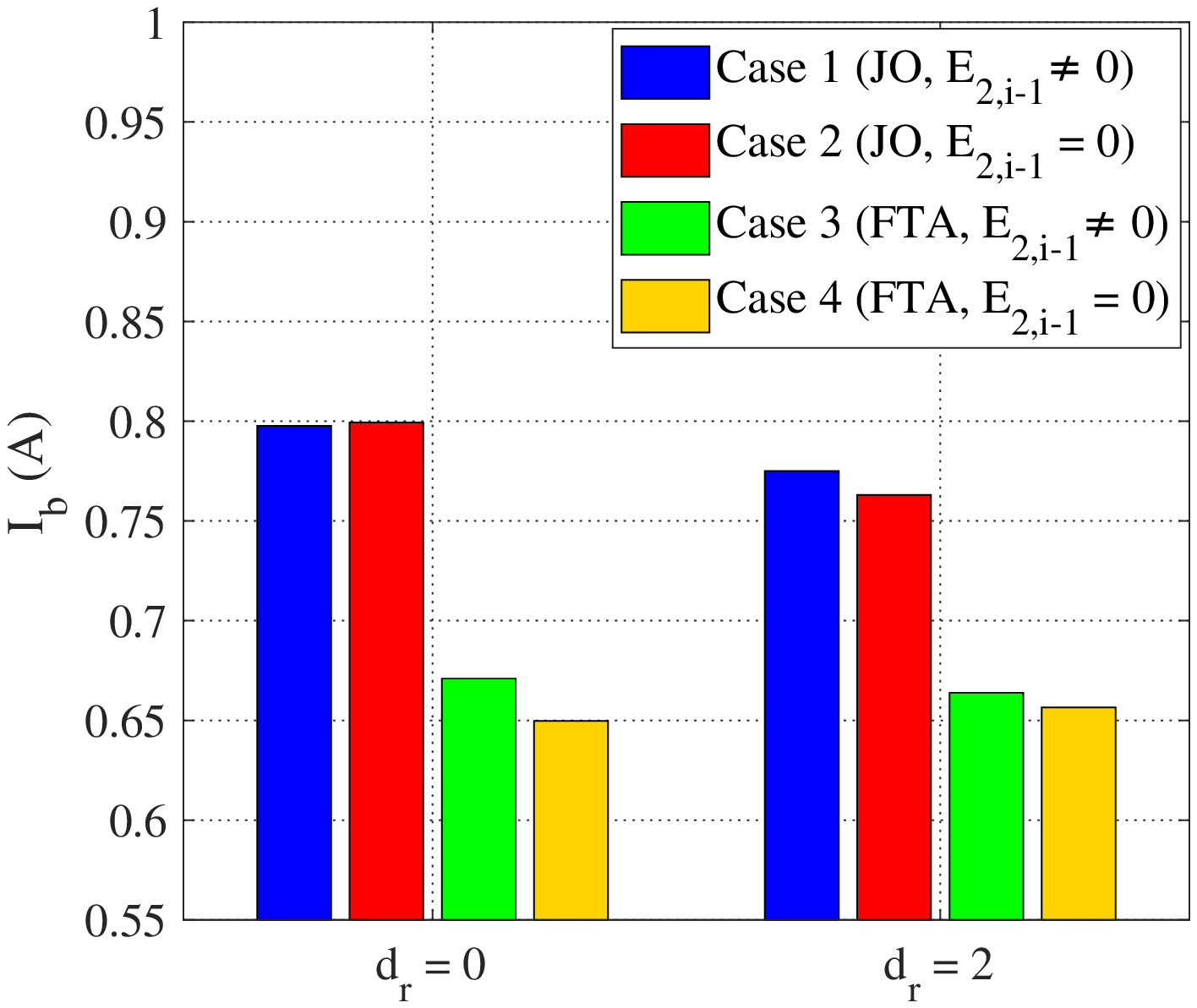} 
\caption{Optimal DC bias}
\label{fig:Ib_random_2G} 
\end{subfigure}
\begin{subfigure}{0.32\linewidth} 
\centering
\includegraphics[trim=0.45cm 0.45cm 0.45cm 0.45cm, clip,width=\linewidth]{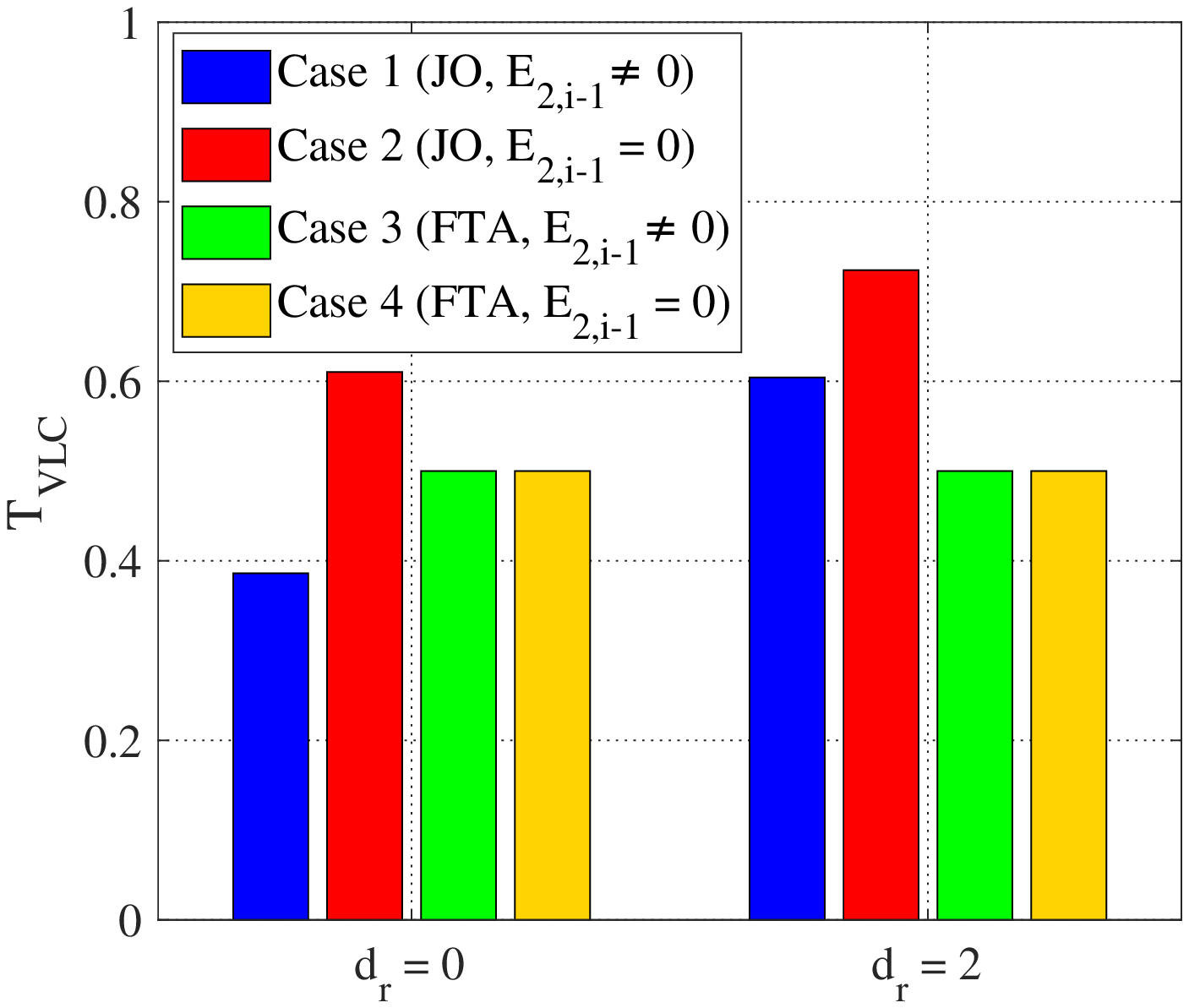} 
\caption{Optimal duration of VLC transmission}\label{fig:T_random_2G}
\end{subfigure} 
\caption{The performance of system under consideration with $f_{\rm c} = 2.4$ GHz when the user node distance follows ${d_{\rm{u}}} \sim \mathcal{U}\left[ {4, 8} \right]$.}
\label{fig:random_locations2G}
\end{figure*}

\begin{itemize}
    \item \textbf{Case 1:} Joint optimization (JO) of $I_{\rm b}$ and $T_{\rm VLC}$ with utilizing the harvested energy from previous transmission block, i.e., $E_{2,\rm{i-1}}\neq 0$; 
    \item \textbf{Case 2:} JO of $I_{\rm b}$ and $T_{\rm VLC}$ without utilizing the harvested energy from previous transmission block, i.e., $E_{2,{\rm i-1}}=0$; 
    \item \textbf{Case 3:} Optimization of $I_{\rm b}$ for fixed time allocation (FTA), i.e., $T_{\rm VLC}=T_{\rm RF}=0.5$, with utilizing the harvested energy from previous transmission block, i.e.,  $E_{2,\rm{i-1}}\neq 0$ (similar to \cite{rakia2016optimal}); 
    \item \textbf{Case 4:} Optimization of $I_{\rm b}$ for FTA, i.e., $T_{\rm VLC}=T_{\rm RF}=0.5$, without utilizing the harvested energy from previous transmission block, i.e., $E_{2,{\rm i-1}}=0$. Similar condition was considered in \cite{peng2021end}.
\end{itemize}

\begin{figure*}[t!]
\centering
\begin{subfigure}{0.32\linewidth} 
\centering
\includegraphics[trim=0.45cm 0.45cm 0.45cm 0.45cm, clip,width=\linewidth]{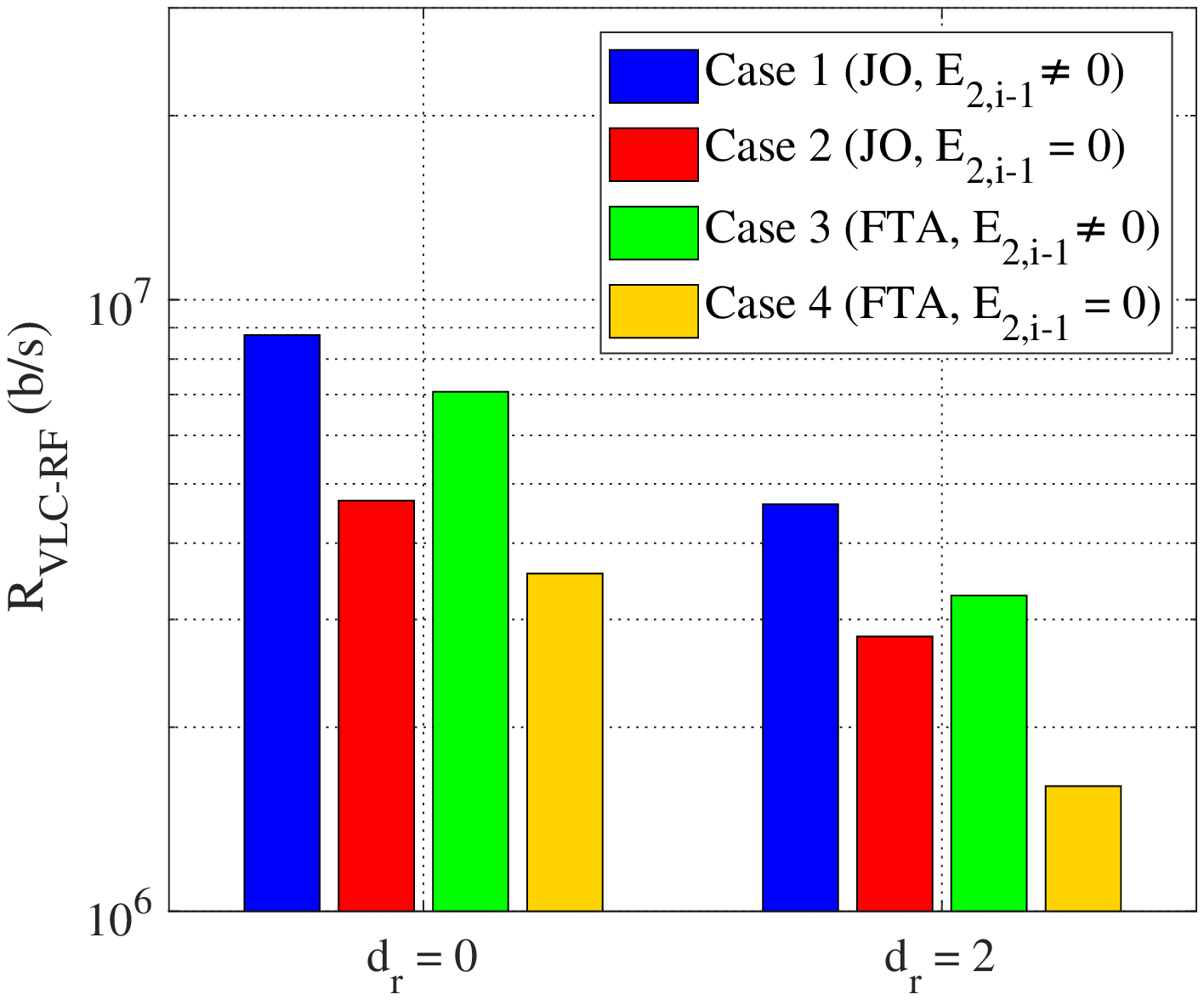} 
\caption{Optimal data rate}\label{fig:rate_random_5G} 
\end{subfigure}
\begin{subfigure}{0.32\linewidth} 
\centering
\includegraphics[trim=0.3cm 0.45cm 0.45cm 0.45cm, clip,width=\linewidth]{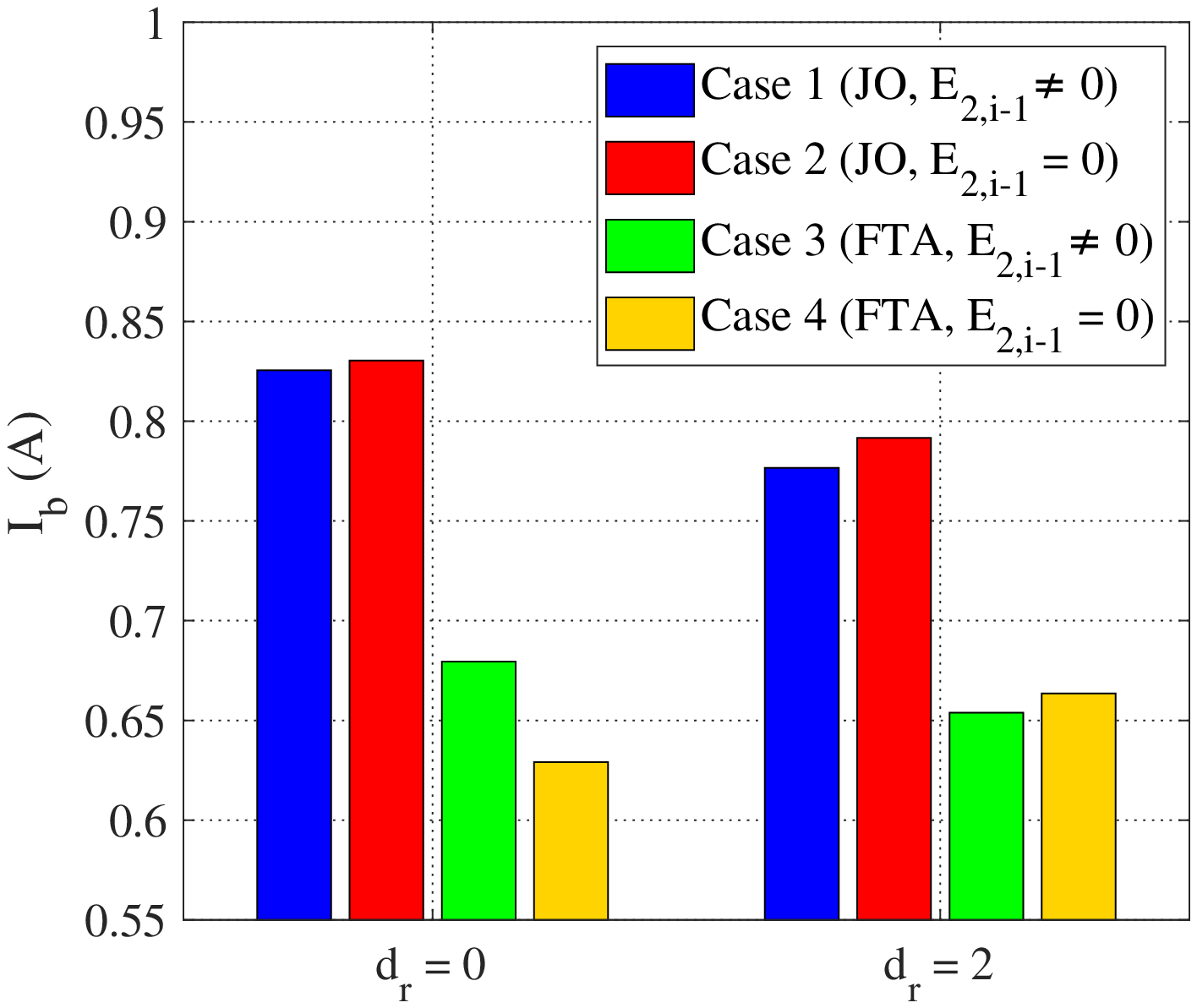} 
\caption{Optimal DC bias}
\label{fig:Ib_random_5G} 
\end{subfigure}
\begin{subfigure}{0.32\linewidth} 
\centering
\includegraphics[trim=0.45cm 0.45cm 0.45cm 0.45cm, clip,width=\linewidth]{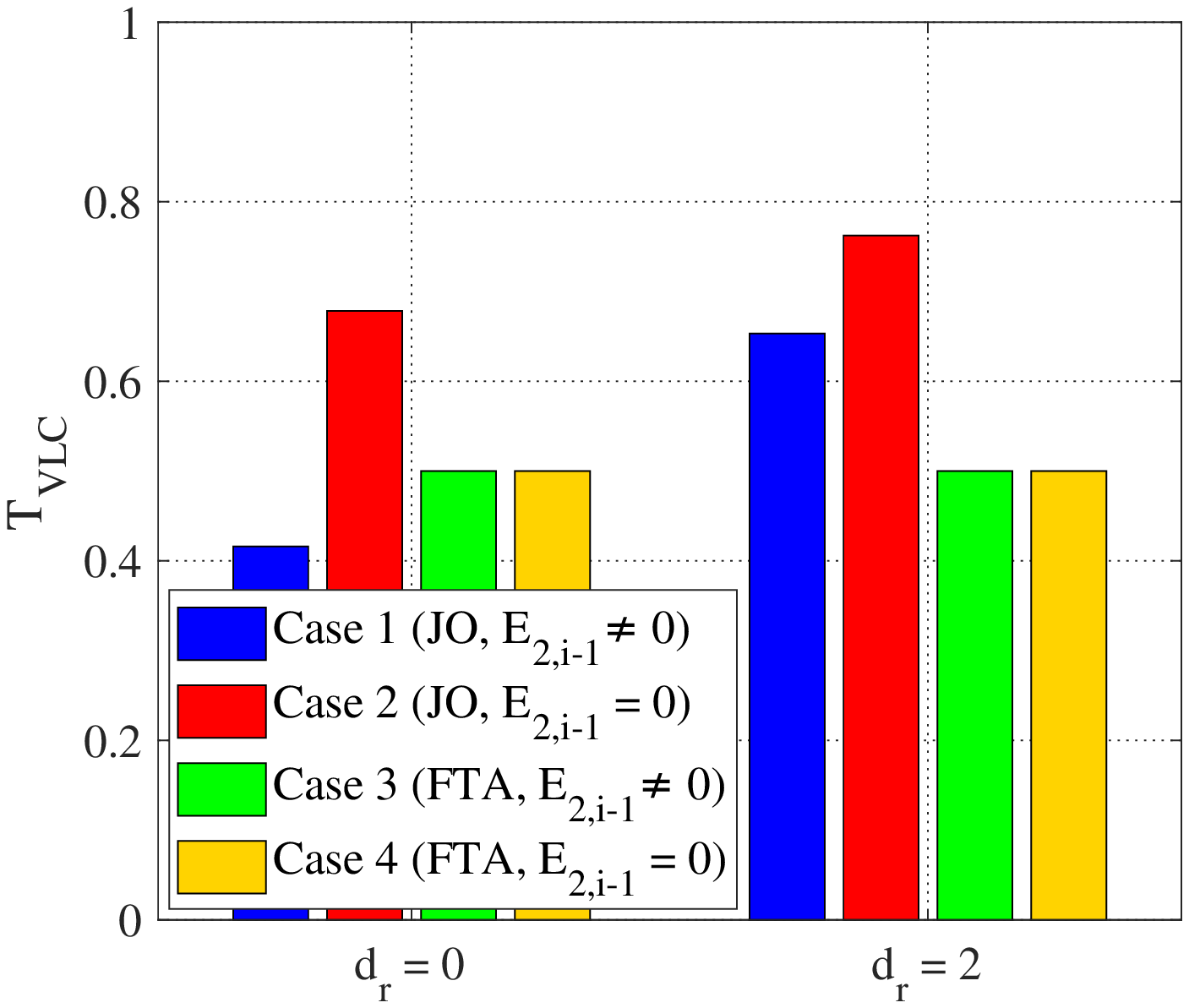} 
\caption{Optimal duration of VLC transmission}\label{fig:T_random_5G}
\end{subfigure} 
\caption{The performance of system under consideration with $f_{\rm c} = 5$ GHz when the user node distance follows ${d_{\rm{u}}} \sim \mathcal{U}\left[ {4, 8} \right]$.}
\label{fig:random_locations5G}
\end{figure*}

Fig.~\ref{fig:rate_diff_user_dis} depicts the performance of optimal data rate when the relay is located at $d_{\rm r} = 0$ and $d_{\rm r} = 2$~m. The user distance varies as $d_{\rm u} \in \{4, 5, 6, 7, 8 \}$. We assume RF frequency as ${f_{\rm{c}} = 2.4}$~GHz. It can be observed that Case 1 and Case 3, where the relay can harvest energy during RF transmission, outperform the other cases. Note the optimal data rate for all cases decreases as the user distance increases. Since the relay node is fixed and the achievable information rate is limited by the smaller information rate between the VLC link and the RF link, one can conclude that the restriction in the system under consideration comes from the RF link. In such energy-deprived regimes, harvesting the energy during the second-hop RF transmission (as proposed in our work) benefits significantly in increasing the bottle-necked data rate.

Fig.~\ref{fig:random_locations2G} illustrates the performance of system under consideration assuming the relay location varies as $d_{\rm r} \in \{0, 2\}$ and the user node distance follow Uniform distribution with ${{d_{\rm{u}}} \sim \mathcal{U}\left[ {4, 8} \right]}$. Here, the RF frequency is assumed as ${f_{\rm{c}} = 2.4}$~GHz. As it can be observed from Fig.~\ref{fig:rate_random_2G}, solving the joint optimization over $I_{\rm{b}}$ and $T_{\rm VLC}$ leads to a higher data rate compared with the other considered cases. In addition, Fig.~\ref{fig:rate_random_2G} clearly shows that harvesting energy during RF transmission (Case 1 and Case 3) results in an increase in optimal data rate. To investigate the reason behind this outperforming, we further depict the optimal DC bias and the time duration assigned for VLC link in~\ref{fig:Ib_random_2G} and~\ref{fig:T_random_2G}, respectively.
The optimal value of DC bias for the joint optimization (regardless of harvesting energy during RF transmission) is higher than the ones obtained when $T_{\rm VLC} = T_{\rm RF} = 0.5$ as shown in Fig.~\ref{fig:Ib_random_2G}. Recalling $c_6^{\prime}$, this leads to a smaller amount of peak amplitude of the input electrical signal which consequently limits the VLC data rate. In addition, it can be observed that the optimal value of $T_{\rm VLC}$ for Case 1 is less than 0.5 which accordingly restrains the data rate of the VLC link (cf. \eqref{eq:2}). Although the difference between the optimal DC bias for Case 1 and Case 2 is practically negligible, optimal $T_{\rm VLC}$ for Case 2 is much higher than Case 1. For Case 2, the power of the relay only relies on the harvested energy during VLC transmission (according to \eqref{eq:4}) which can be increased by allocating more time for this phase. This behavior is because Case 2 ignores harvesting energy during RF transmission and allocating more time for VLC transmission aims to compensate for this effect.

To study the effect of RF frequency, Fig.~\ref{fig:random_locations5G} depicts the performance of system under consideration with $f_{\rm c} = 5$ GHz when the user node distance follows ${d_{\rm{u}}} \sim \mathcal{U}\left[ {4, 8} \right]$ and the relay location varies as $d_{\rm r} \in \{0, 2\}$.
Comparing Fig.~\ref{fig:rate_random_5G} and Fig.~\ref{fig:rate_random_2G} shows that the optimal data rate decreases as the RF frequency increases. According to~\eqref{eq:8}, as RF frequency increases the effect of path loss becomes more pronounced. To further investigate the behavior of the optimization problems, the optimal DC bias and the time duration assigned for the VLC link in Fig.~\ref{fig:Ib_random_5G} and Fig.~\ref{fig:T_random_5G}, respectively. 
For Case 1 and Case 2, the optimal DC bias decreases as the relay distance (i.e., $d_{\rm r}$) increases which leads to a decrease in the VLC data rate. Comparing Fig.~\ref{fig:T_random_5G} and Fig.~\ref{fig:T_random_2G} indicates that the optimal time duration for VLC transmission increases as the RF frequency increases.

\section{Conclusion}\label{sec:conclusion}
In this paper, we proposed an EH hybrid VLC-RF scheme where the relay harvests energy both during VLC transmission and RF communication. With DF relaying assumed, we have formulated the optimization problem to maximize achievable data rate. In addition to the DC bias, we considered the assigned time duration for VLC transmission as a design parameter. Specifically, the joint optimization problem was divided into two sub-problems and solved cyclically. Our first sub-problem involved fixing the time duration for VLC transmission and solving the non-convex DC bias problem using MM approach. In the second sub-problem, we solved the optimization problem for the assigned VLC link time duration using the DC bias obtained in the previous step. The  results have shown that our proposed joint optimization approach achieves a higher data rate compared to solely optimizing the DC bias.

\bibliographystyle{IEEEtran}
\bibliography{references}

\end{document}